\documentclass[%
 reprint, 
 amsmath,amssymb, aps,onecolumn,12pt
]{revtex4-2}

\usepackage{graphicx}
\usepackage{dcolumn}
\usepackage{bm}
\usepackage{braket} 
\usepackage{xcolor}
\usepackage{xr}
\makeatletter
\newcommand*{\addFileDependency}[1]{
  \typeout{(#1)}
  \@addtofilelist{#1}
  \IfFileExists{#1}{}{\typeout{No file #1.}}
}
\makeatother

\DeclareUnicodeCharacter{2212}{-}

\begin{document}
\preprint{APS/123-QED}

\title{
Measurement-Induced Collective Vibrational Quantum Coherence under Spontaneous Raman Scattering in a Liquid
}

\author{Valeria Vento$^\dagger$}
\affiliation{%
 École polytechnique fédérale de Lausanne (EPFL), Institue of Physics, CH-1015 Lausanne, Switzerland
}%
\altaffiliation{These authors contributed equally to this work}
\author{Santiago Tarrago Velez$^\dagger$}
\affiliation{%
 École polytechnique fédérale de Lausanne (EPFL), Institue of Physics, CH-1015 Lausanne, Switzerland
}%
\author{Anna Pogrebna}%
\affiliation{%
  École polytechnique fédérale de Lausanne (EPFL), Institue of Physics, CH-1015 Lausanne, Switzerland
}%
\author{Christophe Galland}
\email{chris.galland@epfl.ch}
\affiliation{%
  École polytechnique fédérale de Lausanne (EPFL), Institue of Physics, CH-1015 Lausanne, Switzerland
}%
\date{\today}
\begin{abstract}

\textbf{Spontaneous vibrational Raman scattering is a ubiquitous form of light-matter interaction whose description necessitates quantization of the electromagnetic field. It is usually considered as an incoherent process because the scattered field lacks any predictable phase relationship with the incoming field. When probing an ensemble of molecules, the question therefore arises: What quantum state should be used to describe the molecular ensemble following spontaneous Stokes scattering? 
We experimentally address this question by measuring time-resolved Stokes--anti-Stokes two-photon coincidences on a molecular liquid consisting of several sub-ensembles with slightly different vibrational frequencies. When spontaneously scattered Stokes photons and subsequent anti-Stokes photons are detected into a single spatiotemporal mode, the observed dynamics is inconsistent with a statistical mixture of individually excited molecules.
Instead, we show that the data are reproduced if Stokes--anti-Stokes correlations are mediated by a collective vibrational quantum, i.e. a coherent superposition of all molecules interacting with light. Our results demonstrate that the degree of coherence in the vibrational state of the liquid is not an intrinsic property of the material system, but rather depends on the optical excitation and detection geometry.}

\end{abstract}

\maketitle

\pagebreak
\textit{Introduction.---} Raman scattering was first reported in 1928 \cite{raman1928} and with the advent of laser sources it has become an essential tool for probing and understanding the vibrational structure of organic and inorganic matter. 
In a majority of experiments, a semi-classical model of light-matter interaction is sufficient to interpret the results of Raman spectroscopy.
For example, the intensity asymmetry between Stokes and anti-Stokes scattering is obtained by quantizing the vibrational modes of each individual molecule.
A full quantum theory of Raman scattering was developped in the 70's and 80's \cite{walls1970a,walls1971,FoerGlaub71,mostowski1981,raymer1981,raymer1985}, and its predictions were tested, e.g., by measuring intensity fluctuations in stimulated Raman scattering \cite{walmsley1983,walmsley1986,raymer1989}.

Following a pioneering work by Walmsley and coworkers in 2011 \cite{lee2011science}, more recent experiments have used time-correlated single photon counting to evidence non-classical intensity correlations between light fields interacting with the same phonon mode via Raman scattering, with potential applications in ultrafast quantum information processing \cite{lee_macroscopic_2012,england2013,england2015,hou2016,fisher2016,fisher2017}, novel forms of spectroscopy \cite{waldermann2008, meiselman_observation_2014}, and the generation of non-classical states of light \cite{tarrago2020}. 
These experimental results have spurred further theoretical developments to understand how the Raman process leads to photonic correlations mediated by a phononic excitation \cite{saraiva2017,Parra-Murillo2016,thapliyal2021, diaz2020}, how the experimental geometry impacts the photon statistics of the Stokes field \cite{shinbrough2020}, and how the coupling of a Raman-active mode to a nanocavity modifies the dynamics of the system \cite{roelli2016,zhang2020,schmidt2021}.

To our knowledge, and despite this recent experimental progress, there has been no direct measurement of the nature of the vibrational quantum state generated in an ensemble of molecules in the liquid phase. 
Since molecule-molecule interactions in a liquid are in general incoherent and only contribute to vibrational relaxation \cite{oxtoby1981}, they cannot generate spatial coherence over mesoscopic length scales -- in contrast to crystalline materials, in which near-field Raman scattering experiments \cite{beams2014,cancado2014,alencar2019} have deduced phonon coherence lengths on the order of tens of nanometers.
Accordingly, most reference texts assume that coherence among different molecules can only be imposed by external driving, e.g. with the beat note between two strong laser fields as in Coherent anti-Stokes Raman scattering (CARS) \cite{schrader2008infrared}. 
Quantum coherence among different molecules following spontaneous Raman scattering in a dense molecular liquid has often been neglected \cite{long2002, leru2008principles}, implicitly assuming that the resulting collective vibrational state is a statistical mixture of individually excited molecules. 
We note that when studying an ensemble of identical molecules, the temporal coherence of the Stokes field \cite{meiselman_observation_2014,sun2021} or the presence of Stokes--anti-Stokes coincidences \cite{bustard2015,Kasperczyk2016} do not provide direct information about the collective coherence possibly existing among the molecules -- which is why the authors from Ref.~\cite{Kasperczyk2016} could describe their experiment in terms of single-molecule scattering events.\\

\begin{figure}[h!]
\centering
\includegraphics[width=0.7\linewidth]{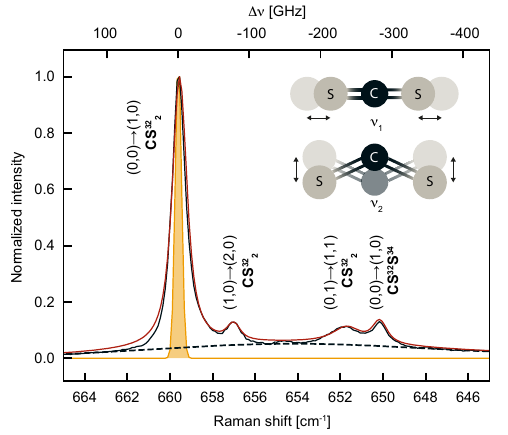}
\caption{Continuous-wave high-resolution Raman spectrum of liquid CS$_2$. The black solid line represents the normalized Stokes intensity measured under $780$~nm  laser excitation. Each peak is associated with a molecular transition of a particular CS$_2$ isotope. In the notation $(n_1^{init.},n_2^{init.})\rightarrow (n_1^{fin.},n_2^{fin.})$,  $n_1$ and $n_2$ are the quantum numbers of the symmetric stretching mode $\nu_1(\Sigma_g^+)$ and the two-fold degenerate bending mode $\nu_2(\Pi_u)$, respectively, as sketched in the inset.
After subtracting a background (dashed line) the spectrum is fitted in frequency space with the sum of four Voigt functions (red line) with fixed Gaussian contribution (orange profile) as described in the text. }
\label{fig:CS2spectrum}
\end{figure}

In this Letter, we demonstrate that spontaneous vibrational Raman scattering in the liquid phase and at room temperature can also induce a collective vibrational state, i.e. a quantum superposition of a macroscopic number of individually excited molecules. 
To this end, we use liquid carbon disulfide (CS$_2$) that naturally contains a few distinct molecular sub-ensembles defined by the initial vibrational state and isotope content of each molecule and having slightly different vibration frequencies (Fig.~\ref{fig:CS2spectrum}). 
We measure time-resolved two-photon Stokes--anti-Stokes correlations mediated by the creation and annihilation of a vibrational quantum in the symmetric stretch mode of the molecules, and observe revivals after several picoseconds. 
These quantum beats are signatures of a coherent superposition where a single vibrational quantum is shared between all molecules involved in Raman scattering and belonging to a spatial mode selected by the measurement geometry. 
Our results are consistent with the numerous works on emissive quantum memories using atomic ensembles \cite{bussieres2013} where collective quantum coherence can be induced by post-selection upon single photon detection -- a key concept underlying the DLCZ proposal for quantum repeaters \cite{duan2001}.
As proposed in the context of cavity optomechanics \cite{flayac2014}, our data are also consistent with the emergence of mode entanglement between molecular sub-ensembles upon single photon detection.

Our experiment, performed in a regime of very low Raman scattering probability (about one per hundred pulses), departs from the vast body of literature reporting beat notes under stimulated Raman scattering and other nonlinear spectroscopic data, where classical coherence is imposed by the nonlinear interaction with the excitation beams \cite{laubereau_collective_1976,pestov2007, konarska_dynamics_2016}. 
Our results are also distinct from the quantum beats observed when different energy levels of a same molecule are coherently excited \cite{bitto1990,walmsley1988,carter2000}, since each single molecule within our ensemble vibrates at a single frequency, and only the collective excitation displays quantum beats. \\

\textit{Sample characterization.---} Figure~\ref{fig:CS2spectrum} shows the Raman spectrum of liquid CS$_2$ (anhydrous, $\geq$99 \%, Sigma Aldrich) at room temperature, zooming in on the Raman shift of the symmetric stretch mode $\nu_1$, acquired under $780$~nm continuous-wave excitation with a high-resolution Shamrock SR-750 spectrometer equipped with a Newton 940 camera (Andor) and a $1800$~l/mm $400$~nm blaze grating. 
Note that CS$_2$ has no electronic transition at visible or near-infrared frequencies so that all measurements reported here correspond to far off-resonance excitation.
The vibrational transitions associated to the four main peaks in Fig.~\ref{fig:CS2spectrum} are assigned following Refs.~\cite{cox_properties_1979, stoicheff_high_1958, plyler_infrared_1947}, and are labelled according to the initial (at thermal equilibrium) and final (after Stokes scattering) occupation numbers of the stretching ($\nu_1$) and bending ($\nu_2$) vibrational modes. 

We recognize the pure $\nu_1$ bands of the two dominant isotopes CS$_2^{32}$ and CS$^{32}$S$^{34}$ at 659.6 cm$^{-1}$ and 650.1 cm$^{-1}$, respectively, and two hot bands of CS$_2^{32}$ at 657 cm$^{-1}$ and 651.7 cm$^{-1}$. 
In the following, these four dominant vibrational transitions are simply labeled with $i= 1,2,3,4$ in order of decreasing Raman shift. 
The pure $\nu_1$ band of CS$^{32}$S$^{33}$ (barely distinguishable around 655 cm$^{-1}$) and other hot bands of the three isotopes CS$_2^{32}$, CS$^{32}$S$^{33}$ and CS$^{32}$S$^{34}$ contribute to the background used in our fit and marked as dashed lines. 

After background subtraction we fitted the Raman spectrum in frequency space with the sum of four Voigt functions: for each peak, we assume that the Raman line shape is Lorentzian, while the Gaussian contribution is fixed by the instrument response function. 
The latter is estimated by measuring the spectrum of the attenuated excitation laser, which is well fit with a Gaussian of $\sim 0.02$~nm FWHM (orange shaded peak), in agreement with the nominal system resolution. 
The free fit parameters are the central frequencies of the four Raman peaks and the FWHMs $\Delta\nu_i$ of three modes ($i= 1,2,3$) -- the pure $\nu_1$ vibrations of both isotopes are supposed to have the same linewidth, $\Delta\nu_1=\Delta\nu_4$. 
The relative intensities among all peaks are fixed by temperature and isotope abundance following Ref.~\cite{cox_properties_1979}.
From the Raman shifts, we obtain the four vibrational frequencies: $20.04$, $19.96$, $19.81$ and $19.76$~THz. 
From the Lorentzian width, we infer the effective coherence time $T_2 = (\pi\Delta\nu)^{-1} $ of each vibration in our measurement geometry: 15.71, 18.38, 6.77 and 15.71~ps respectively.\\


\begin{figure}[h!]
\centering
\includegraphics[width=0.7\linewidth]{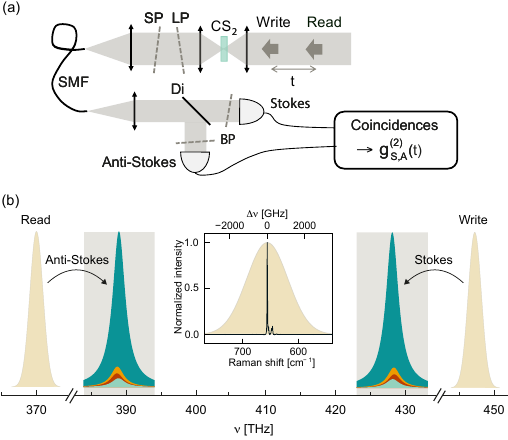}
\caption{\textbf{(a)} Sketch of the experimental setup. Liquid CS$_2$ is held between two objective lenses (numerical aperture: 0.8) in a quartz cuvette sealed with parafilm with $\sim 0.2$~mm wall thickness and $\sim 1$~mm optical path.
Fiber-coupled single photon avalanche photodiodes are connected to a coincidence counter to measure $g^{(2)}_{S,A}$.
SP: shortpass; LP: longpass; BP:bandpass; Di: dichroic; SMF: single mode fiber.
\textbf{(b)} Frequency-domain schematic of experimentally relevant optical fields. 
Only the Stokes signal from the write pulse and the anti-Stokes signal from the read pulse are transmitted through the filters (grey area) to their respective detectors. Each Raman peak inherits the linewidth of the excitation pulse ($\sim 200$~fs duration) so that adjacent vibrational modes (Fig.~\ref{fig:CS2spectrum}) are no longer distinguishable. Inset: comparison between the excitation pulse width and the continuous-wave Raman spectrum. }
\label{fig:Fig_experiment}
\end{figure}

\textit{Experimental Results.---}
We use the technique introduced in \cite{anderson_two-color_2018}; a simplified description of the experimental setup is provided in Fig.~\ref{fig:Fig_experiment}. 
A first near-infrared laser pulse ($\sim 200$ fs pulse duration, $80$~MHz repetition rate) generates a two-mode photon-vibration squeezed state via spontaneous Stokes scattering. 
The power is adjusted to generate less than $10^{-2}$ Stokes photon per pulse within the spatial mode selected by the collection fiber, well below the onset of any stimulated Raman process, {as confirmed by the linear power-dependence of both Stokes and anti-Stokes scattered intensities (see Supplementary Material)}. 
After this \textit{write} pulse, a \textit{read} pulse centered at a different wavelength is used to probe the vibrational mode through anti-Stokes scattering after a variable time delay $t$.
The Stokes--anti-Stokes correlation function $g^{(2)}_{S,A}(t)$ reflects the decay of the vibrational excitation heralded by the detection of a Stokes photon \cite{galland2014,riedinger2016}. 
In the limit of low scattering probability and absent any background emission nor noise, the strength of vibration-mediated correlation is upper-bounded by $g^{(2)}_{S,A}< 1 + 1/n_{th} \simeq 26$, where the thermal occupancy of the vibrational mode is $n_{th} \simeq 0.04$ in our system at room-temperature.

The sample is studied in transmission to fulfill momentum conservation, where the Stokes -- anti-Stokes process is mediated by a collective vibration with vanishing momentum. 
The Raman signal is collected into a single-mode optical fiber whose back-propagated image overlaps with the focused laser beams to define a single spatial mode inside the sample. 
After spatial filtering through the fiber, the Stokes and anti-Stokes photons from the first and second pulses, respectively, are separated based on their non-overlapping spectra and individually detected.\\

\begin{figure}[ht]
\centering
\includegraphics[width=0.7\linewidth]{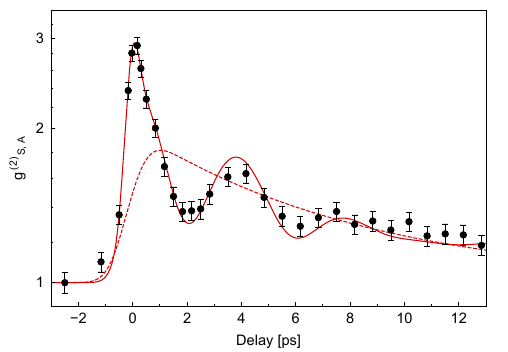}
\caption{Write-read delay dependence of the measured normalised Stokes--anti-Stokes correlations (full circles) and model prediction (solid red line). 
Only Raman photons co-polarised with their respective laser pulses were selected.
Free fitting parameters entering the model curve are a scaling amplitude $A_\text{tot}=1.67$ for the oscillations and a rising time $T_\text{rise}=0.66$ ps. To account for electronic four-wave mixing we add a Gaussian peak centered at zero delay of amplitude  $A_g=1.06$. 
The dotted red line represents the multi-exponential decay that would result from a statistical mixture of single vibrating molecules. 
}
\label{fig:g2_osc}
\end{figure}

Figure~\ref{fig:g2_osc} shows $g^{(2)}_{S,A}(t)$ plotted as a function of time delay between Stokes and anti-Stokes processes. 
When write and read pulses temporally overlap, virtual electronic processes contribute to four-wave mixing (FWM) and generate photon pairs at any frequencies satisfying energy conservation (phase matching is highly relaxed in our strongly focused geometry). 
We used crossed-polarised laser pulses to minimize the relative contribution of electronic FWM to the overall signal. 
Beyond $t \simeq 1$~ps, instantaneous contributions to FWM have vanished and $g^{(2)}_{S,A}(t)>1$ is a signature of intensity correlations between Stokes and anti-Stokes fields mediated by molecular vibrations.
The red dashed curve is a tentative fit with a multi-exponential decay whose relative amplitudes and time constants are extracted from Fig.~\ref{fig:CS2spectrum}. 
This decay would result from describing the vibrational state as a statistical mixture with one molecule excited at random for each coincidence count. 
While this model matches the overall damping of correlations, it fails to capture the clear oscillations in the value of $g^{(2)}_{S,A}(t)$. \\

\textit{Discussion.--- }
We note that oscillations resulting from the excitation of vibrational modes in different isotopes of CCl$_4$ \cite{wu_isotope_1938} were observed using ultrafast stimulated Raman scattering in Ref.~\cite{laubereau_collective_1976, konarska_dynamics_2016}. 
However such experiments are well accounted for by a semi-classical model, in which the stimulated Stokes and anti-Stokes fields are classical. In the case of stimulated Raman scattering, the beating between the two pump laser fields is tuned resonant with the molecular vibration of interest, thereby driving a coherent collective vibration and resulting in a coherently oscillating Raman polarisation in the sample, all of which behave as classical variables.

In contrast, spontaneous Raman scattering and single photon counting demand a quantum description, in which the post-selected vibrational state naturally appears as a quantum coherent superposition involving all molecules coupled to the light field.
Such a quantum model including experimental noise sources is detailed in the Supplementary Material and solved numerically; here we prefer to present instead a simplified analytical model that captures the essence of the phenomenon and does reproduce the observed quantum beats (cf. red solid line in Fig~\ref{fig:g2_osc}).\\ 

We estimate that about $N\sim10^{10}$ randomly moving molecules occupy the focal volume. If we consider them individually, the Raman interaction can be modeled as a product of $N$ two-mode squeezing unitary operators $\hat{U}_k$ acting on the vacuum state of the Stokes photon ($S$) and the vibration ($v$) 
\cite{galland2014, tarrago2020} (we neglect thermal excitation for the sake of simplicity). 
Assuming the same Raman cross-section for all the molecules and uniform field intensity within the focal volume, and ignoring corrections from time-ordering of operators \cite{christ2013,quesada2014} as valid in the low-gain regime, the state of the system after a short interaction is 
\begin{equation}
\begin{split}
    \ket{\psi} &=\hat{U}\ket{\text{vac}} \\ 
    &= \bigotimes_{k=1}^N \hat{U}_k\ket{\text{vac}} = \bigotimes_{k=1}^N e^{p \hat{A}_{S,k} \hat{A}_{v,k}-h.c.}\ket{\text{vac}} \\ 
    &= \ket{\text{vac}} + \sqrt{p}\sum_{k=1}^N \ket{1}_{S,k} \ket{1}_{v,k} +  O(p)
\end{split}
\end{equation}
where $p\simeq 10^{-2}$ in our experimental conditions (low gain, spontaneous regime). $\hat{A}_{S,k}$ and $\hat{A}_{v,k}$ are the annihilation operators of Stokes photon and of the vibration, respectively, and $\ket{1}_{S,k} \ket{1}_{v,k} = \hat{A}_{S,k}\hat{A}_{v,k} \ket{\text{vac}}$. 

After the detection of a Stokes photon, the post-selected vibrational state is given by 
\begin{equation}\label{eq:postsel}
    \rho_\text{ps}^{(j)}=\frac{\text{Tr}_v\left(\hat{K}_j\rho\hat{K}_j^\dagger\right)}{\text{Tr}\left(\rho\hat{K}_j^\dagger\hat{K}_j\right)}
\end{equation}
where $\rho=\ket{\psi}\bra{\psi}$ and $\hat{K}_j$ 
are Kraus operators relative to a particular outcome $j$. A proper choice of Kraus operator is therefore key to model the experimentally observed vibrational state.
Since the inverse duration of the laser pulses is much broader than the differences in vibration frequency between distinct molecules and since we collect the forward-scattered Raman signal through a single-mode fiber, information about which molecule is excited is erased, both spectrally and spatially. In this limit, the detection process can be described by a single Kraus operator $\hat{K}=\ket{\text{vac}}\bra{\phi}$, where $\ket{\phi}=\frac{1}{\sqrt{N}}\sum_{k=1}^{N}\ket{1}_{S,k}$. 
Here, we have idealized the collective bright state as having zero momentum (only real coefficients). 

Since the denominator of Eq.~(\ref{eq:postsel}) is equal to $p$, it results $\rho_\text{ps}(t=0)\equiv \rho_\text{ps}^{(1)}= \ket{\psi_\text{ps}(0)}\bra{\psi_\text{ps}(0)}$ with
\begin{equation}
    \ket{\psi_\text{ps}(0)} = \frac{1}{\sqrt{N}}\sum_{k=1}^N\ket{1}_{v,k}.
\end{equation}

From the cw Raman spectrum, we found four collective vibrational modes, which correspond to four sub-ensembles of molecules $i=1,2,3,4$ with frequencies $\Omega_i$ (in rad/s). Therefore, we can rewrite the post-selected state as 
\begin{equation}
    \ket{\psi_\text{ps}(0)} = 
    \sum_{i=1}^4\beta_i\ket{\chi_i}_v
\end{equation}
where
\begin{equation}
    \ket{\chi_i}_v = \frac{1}{\sqrt{N_i}}\sum_{k=1}^{N_i} \ket{1}_{v,k}
\end{equation}
is the state of a single collective excitation of the subensemble $i$, and $\beta_i^2=N_i/N$ is the fraction of molecules in subensemble $i$. This quantity is equal to the fractional intensity of the corresponding cw Raman peak as extracted from Ref. \cite{cox_properties_1979}, i.e. $\beta_i^2=0.70$, $0.03$, $0.21$ and $0.06$. Therefore, the post-selected vibrational state at time $t$ after Stokes scattering is given by
\begin{equation}
    \ket{\psi_\text{ps}(t)} = 
    \sum_{i=1}^4\beta_ie^{-t/T_{2,i}-i\Omega_it}\ket{\chi_i}_v
\end{equation}
 where the coherence times $T_{2,i}$ are extracted from the spectrum in Fig.~\ref{fig:CS2spectrum}.
 
The probability of generating an anti-Stokes photon at time $t$ from this post-selected vibrational state is proportional to
\begin{equation}
P_A(t)\propto \left|\braket{\psi_\text{ps}(0)|\psi_\text{ps}(t)}\right|^2=\left|\sum_{i=1}^4\xi_i(t)\right|^2
\end{equation}
where $\xi_i(t)=\beta_i^2e^{-t/T_{2,i}-i\Omega_it}$. 
Here, we used the fact that the anti-Stokes detection geometry selects the same spatial mode as excited in Stokes scattering.
We see that the Stokes--anti-Stokes correlations feature interference between the complex amplitudes $\xi_i(t)$, such that 
\begin{equation}
    g^{(2)}_\text{model}(t) - 1 = A_\text{tot} \left|\sum_{i=1}^4 \xi_i(t)\right|^2.
\end{equation}
The complete fit function has the following expression:
\begin{equation}
    F_\text{model}=g^{(2)}_\text{model}*f_\text{rise}+G_\text{FWM}.
\end{equation}
where we convolute with a rise function and add a Gaussian function centered at zero delay that accounts for electronic FWM (which has vanishing memory compared to vibrational contributions). With this model we are able to fit the oscillatory decay of the heralded anti-Stokes intensity using only one scaling amplitude $A_\text{tot}$ as free parameter (plus a rise time $T_\text{rise}$ and the amplitude $A_g$ of the electronic FWM to fit the data close to zero delay), see Fig.~\ref{fig:g2_osc}. 
{The value of $T_\text{rise}$ is longer than the nominal pulse duration due to dispersion from all the optical elements before the sample, which is not compensated for.}
The scaling factor $A_\text{tot}$ emerges naturally from the full quantum model solved numerically in the SM, once accounting for main sources of noise and inefficiencies. 

\begin{figure}[h!]
\centering
\includegraphics[width=0.7\linewidth]{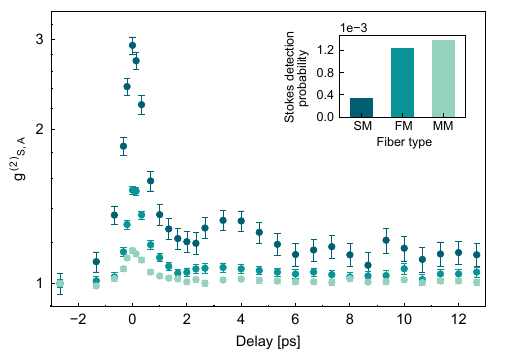}
\caption{Measured  $g^{(2)}_{S,A}(t)$ using different optical fibers in collection path (see Fig.~\ref{fig:Fig_experiment}a). From darker to lighter color: $3.5$-$\mu$m core single-mode fiber (SM), $9$-$\mu$m core telecom fiber (FM, few-mode), and $50$-$\mu$m core multimode (MM) fiber. In the inset, we show the Stokes detection probability for the different fibers. 
}
\label{fig:g2_fibers}
\end{figure}

While the quality of the fit provides strong support for our model, similar quantum beats would also result from coherence between energy levels of a single molecule \cite{bitto1990,walmsley1988,carter2000}. 
In such a case, the exact detection geometry should have no impact on the observation, and in particular it would not be essential to post-select a single spatial mode to observe quantum beats. 
To test this hypothesis we repeat the same measurement using increasing fiber core sizes in the collection path, and therefore increasing number of spatial modes contributing to the total signal, with results shown in Fig.~\ref{fig:g2_fibers}.
The loss of visibility when measuring multiple optical modes provides further evidence that the observations are due to the spatial coherence generated through the experimental geometry and measurement post-selection. 
While a complete model for the multi-mode case is left for future work, we propose to interpret this result in light of information erasure: as the fiber core size increases, more and more information becomes available about the spatial origin of the signal, affecting the coherence of the post-selected state. 

\textit{Conclusions.---} To summarize, we presented a time domain measurement of spectrally resolved Stokes--anti-Stokes two-photon correlations on liquid CS$_2$ at room temperature, in the regime of spontaneous Raman scattering (Stokes scattering probability $\sim$1\%). 
Upon post-selection of events where both Stokes and anti-Stokes photons are collected through the same single-mode optical fiber, we observe quantum beats that are perfectly consistent with a macroscopic quantum superposition of molecules sharing a single quantum of vibration.  
In particular, our experimental results are incompatible with the picture according to which spontaneous Raman scattering from a large ensemble is always the incoherent sum (statistical mixture) of single-molecule scattering events.
Our experiment nourishes the debate about the relation between optical coherence and quantum coherence \cite{molmer1997optical,bartlett2006} and entanglement \cite{karnieli2021}. 
It questions whether optical coherent states are necessary to explain various forms of coherent Raman spectroscopy, for we do not stimulate the Raman process and yet observe coherent oscillations, as if we did.
In the future, our photon counting approach can be adapted to probe inter- or intramolecular vibrational entanglement in more complex systems, as well as excitonic and vibrational polariton dynamics \cite{dorfman2018}. 
We also envision extensions of our work to probe how Raman scattering is affected by collective excitonic \cite{galego2015,spano2015,herrera2016,avramenko2019} or vibrational \cite{shalabney2015,delPino2015,delPino2016,strashko2016,roelli2020,hughes2021} strong coupling to a cavity, with implications for polariton chemistry \cite{ribeiro2018}.

\subsection*{Acknowledgements}
The authors thank Vivishek Sudhir for insightful discussions and valuable comments.  
This work has received funding from the Swiss National Science Foundation (SNSF) (projects no. 170684 and 198898) and the European Research Council’s (ERC) Horizon 2020 research and innovation programme (grant agreement no. 820196)
A. P. acknowledges funding from the European Union's Horizon 2020 research and innovation programme under the Marie Skłodowska-Curie grant agreement N$^{\circ}$ 754462.

\subsection*{Data and code availability statement} 
Data and codes that support the findings of this study have been deposited in a Zenodo repository [DOI to be included here]

\bibliography{biblio}

\newpage

\renewcommand{\figurename}{\textbf{Supplementary Figure}}
\renewcommand{\tablename}{\textbf{Supplementary Table}}
\renewcommand{\thefigure}{S\arabic{figure}}

\section*{Supplementary Material}

\textit{Full quantum model.---} 
The four collective vibrational modes of interest in our experiment are described by the annihilation operators $\hat{b}_1$, $\hat{b}_2$, $\hat{b}_3$, $\hat{b}_4$. 
Under sufficiently short excitation pulse and single spatial mode filtering, all four vibrational modes effectively couple to the same Stokes and anti-Stokes photon fields, described by the annihilation operators $\hat{a}^{x}_{S}$, $\hat{a}^{x}_{A}$, where $x = w,r$ for the write and read pulses, respectively. The Raman interaction is modeled by the Hamiltonian \cite{FoerGlaub71}
\begin{equation}\label{completeH}
\begin{split}
  \hat{H}_{I}^x = \hbar \alpha^x \left[ \lambda^x_{S} (\hat{a}^x_{S})^\dagger (\beta_1 \hat{b}_1^\dagger + e^{-i \theta^x_{2,S}} \beta_2 \hat{b}_2^\dagger \right. \\ \left. + e^{-i \theta^x_{3,S}} \beta_3 \hat{b}_3^\dagger + e^{-i \theta^x_{4,S}} \beta_4 \hat{b}_4^\dagger )\right.\\
  +\left. \lambda^x_{A} (\hat{a}^x_{A})^\dagger (\beta_1 \hat{b}_1 + e^{-i \theta^x_{2,A}} \beta_2 \hat{b}_2 \right.\\ \left.+ e^{-i \theta^x_{3,A}} \beta_3 \hat{b}_3 + e^{-i \theta^x_{4,A}} \beta_4 \hat{b}_4)  \right] + h.c.
\end{split}
\end{equation}
where the laser is modeled by a coherent field of amplitude $\alpha_x (t)$ ($x = w,r$) with Gaussian envelope centered at time $ t_{0x} $ of width $\sigma_x$,
\begin{equation}
  \alpha^x (t) =  A_x \exp \left( \frac{(t-t_{0x})^2}{2 \sigma_x^2} \right) \exp \left( -i \omega_x t \right).
\end{equation}
In eq.~(\ref{completeH}) $\lambda^x_{S,A}$ determine the coupling strengths to the Stokes and anti-Stokes modes, and the $\beta_i$'s are real numbers satisfying $\sum_i \beta_i^2 =1$ where $\beta_i^2$ control the relative intensity of the Raman peak corresponding to vibrational mode $i$, experimentally determined by the relative isotopic abundance and the temperature.

Since we use spectral filtering and post-selection to ignore events where an anti-Stokes (resp. Stokes) photon is emitted during the write (resp. read) pulse we can simplify the interaction model to
\begin{align}
  \hat{H}_{I}^w = &\hbar \lambda^w_{S} \alpha^w (\hat{a}^w_{S})^\dagger (\beta_1 \hat{b}_1^\dagger + e^{-i \theta^w_{2,S}} \beta_2 \hat{b}_2^\dagger  \\ &+ e^{-i \theta^w_{3,S}} \beta_3 \hat{b}_3^\dagger + e^{-i \theta^w_{4,S}} \beta_4 \hat{b}_4^\dagger ) + h.c.\\
  \hat{H}_{I}^r = &\hbar \lambda^r_{A} \alpha^r (\hat{a}^r_{A})^\dagger (\beta_1 \hat{b}_1 + e^{-i \theta^r_{2,A}} \beta_2 \hat{b}_2\\ &+ e^{-i \theta^r_{3,A}} \beta_3 \hat{b}_3 + e^{-i \theta^r_{4,A}} \beta_4 \hat{b}_4 ) + h.c. 
\end{align}
(Note that the ignored terms contribute to uncorrelated noise photons generated via higher order Raman interactions during a single pulse \cite{Parra-Murillo2016}, which is accounted for below.)

We have explicitly written the phase differences $\theta^w_{i,S}$, resp. $\theta^r_{i,A}$, appearing between the three vibrational modes ($i=2,3,4$) during Stokes, resp. anti-Stokes, scattering, taking the global phase such that $\theta^{w,r}_{1}=0$. The experiment is however only sensitive to the sums $\theta_i=\theta^w_{i,S}+\theta^r_{i,A}$. Interestingly, we find excellent fit with the data for $\theta_i=0$ for $i=2,3,4$, which we tentatively relate to the very large detuning between the laser pulses and any electronic resonance of CS$_2$.
In the following we shorten the notation for the annihilation operators to $\hat{a}^w_{S} \equiv \hat{a}_{S}$ and $\hat{a}^r_{A} \equiv \hat{a}_{A}$.

We include an additional $\chi^{(3)}$ nonlinear interaction term that allows for the direct interaction between the write and read pulses, leading to the creation of photon pairs at the frequencies of the Stokes and anti-Stokes emission (FWM process):
\begin{equation}
  \hat{H}_{I}^{(3)} = \hbar \lambda^{(3)} \alpha^{w} \alpha^{r} \hat{a}_{S} \hat{a}_{A} +h.c.
\end{equation}
In the frame rotating with the central frequency $\omega_0 = \frac{\omega_w+\omega_r}{2}$ we obtain the effective Hamiltonian
\begin{equation}
\begin{split}
  \hat{H} = \hbar\Omega_{1}\hat{b}^\dagger_1\hat{b}_1 + \hbar\Omega_{2}\hat{b}^\dagger_2\hat{b}_2 + \hbar\Omega_{3}\hat{b}^\dagger_3\hat{b}_3 + \hbar\Omega_{4}\hat{b}^\dagger_4\hat{b}_4 \\ + \hbar\Delta_{S}\hat{a}^\dagger_S\hat{a}_S +  \hbar\Delta_{A}\hat{a}^\dagger_{A}\hat{a}_{A} \\+ \hat{H}_{I}^w + \hat{H}_{I}^r + \hat{H}_{I}^{(3)}
\end{split}
\end{equation}
where $\Delta_{S,A} = \omega_{S,A} - \omega_0$.
To account for dissipation we use the master equation approach, which includes coupling of the phonon modes to a thermal reservoir at room temperature,  as described by the collapse operators
\begin{equation}
\begin{split}
    &\hat{C}_{bi-} = \sqrt{\kappa_i (1+n_{th})} \hat{b}_i \\
    &\hat{C}_{bi+} = \sqrt{\kappa_i n_{th}} \hat{b}_i^\dagger
\end{split}
\end{equation}
where $i=1,2,3,4$ and $\kappa_i$ is related to the decay rate of the phonon modes by $\tau_{ph_i} = 1/\kappa_i$.
The temporal evolution of the density matrix is computed numerically using QuTiP, an open-source library used for simulating quantum systems in Python \cite{johansson2012, johansson2013}.

\textit{Coincidence counts under non-ideal conditions.---} To account for noise and losses in the experiment we use the operator 
introduced in \cite{sekatski2012} to model the photon detection probability
\begin{equation}
  \hat{D}_{X} = 1 - (1-p^{dc}_{X})(1-\eta_{X})^{\hat{a}^\dagger_{X} \hat{a}_{X}}
\end{equation}
where $X=S,A$ for the Stokes and anti-Stokes detection channels, respectively. The dark count probability (per detection time window) is $p^{dc}_{X}$ while $\eta_{X}$ is the detection efficiency.
The experimental value of the normalised Stokes -- anti-Stokes coincidence rate $g^{(2)}_{S,A}$ is then calculated (for several values of $t = t_{0r} -t_{0w}$) as
\begin{equation}
  g^{(2)}_{S,A} = \frac{\left< \hat{D}_{S} \hat{D}_{A} \right>}{\left< \hat{D}_{S} \right> \left< \hat{D}_{A} \right>}. 
\end{equation}

\textit{Comparison with the analytical model.---} 
In order to corroborate the quality of the analytical model that we presented in the main text, we used the full quantum model with the analytically-extracted parameters to simulate our experimental results. Therefore, we set $\theta_2 = \theta_3 = \theta_4 = 0 $; $\Omega_i/(2\pi) \simeq 20.04$, $19.96$, $19.81$ and $19.76$~THz and $\tau_{ph_i} = T_i/2$ (see Ref.\cite{velez_preparation_2019}) with $T_i \simeq 15.71$, $18.38$,  $6.77$ and $15.71$~ps as extracted from the cw spectrum, and the weights $\beta_i^2 \simeq 0.70$, $0.03$, $0.21$ and $0.06$.

The detection efficiency of our setup is estimated to be $\eta_S \approx \eta_{A} \approx 10\%$. We fix the values of the parameters $A_x$ and $\lambda^x_{S,A}$ by considering the Stokes and anti-Stokes detection rates.  As the pulse amplitudes $A_x$ always appear in factor with the coupling rates $\lambda^x$, we introduce $\Lambda^x_{S,A} = A_x \lambda^x_{S,A}$ whose values are chosen to reproduce the measured single-detector count rate when only the write or read pulse is propagating through the sample. We find $\Lambda^w_{S} = 0.102$ and $\Lambda^r_{A} = 0.131$, which recover the measured detection probabilities $p_S \approx 2.5 \times 10^{-4}$, $p_{A} \approx 1.6 \times 10^{-5}$.
The dark count probabilities are estimated to be $p^{dc}_S \approx 2 \times 10^{-4}$ and $p^{dc}_A \approx 1 \times 10^{-5}$ and include detector noise plus background emission noise. 
The coupling rate accounting for all the FWM processes at zero delay has been set to $\Lambda_{FWM}= A_wA_r\lambda^{(3)} = 0.0094$ in order to reproduce the experimental results. 
\begin{figure}[]
\centering
\includegraphics[width=0.6\linewidth]{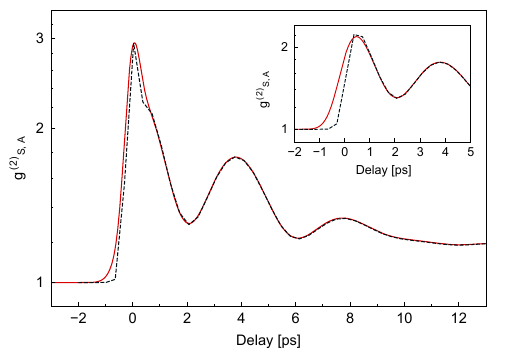}
\caption{Comparison between the full quantum model (dashed black line) and the analytical model (solid red line). In the inset, the two models are shown with FWM hamiltonian $\hat{H}_{I}^{(3)}=0$.
}
\label{fig:comparison}
\end{figure}

Figure~\ref{fig:comparison} shows the match between the two models, where the discrepancy around zero delay is due additional dispersion in the optics that increases the rise time (not considered in the quantum model).

\begin{figure}[ht]
\centering
\includegraphics[width=0.6\linewidth]{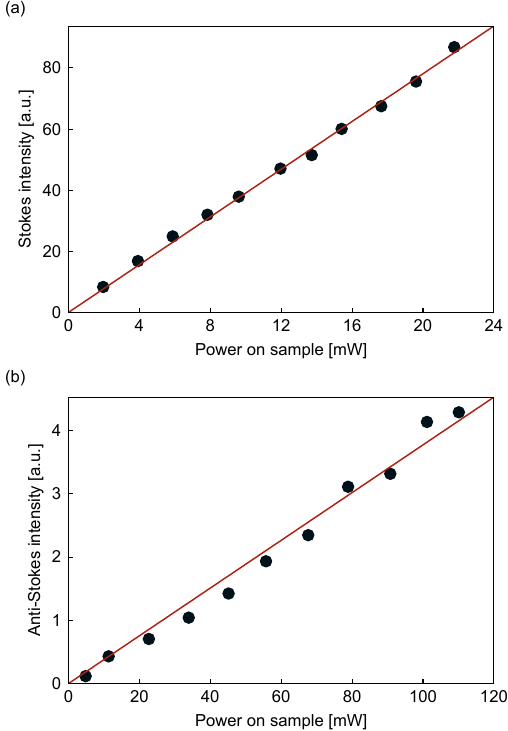}
\caption{\textbf{(a)} Stokes intensity as a function of the write power on the sample. \textbf{(b)} Anti-Stokes intensity as a function of the read power on the sample. The counts are integrated over the relevant range of the Raman spectrum corresponding to the correlation measurements. The red lines are linear regressions.}  
\label{fig:power}
\end{figure}
\textit{Power dependence.---}
For each measurement, the time-averaged write power on the sample is in the range [$5$-$7$]mW, while the time-averaged read power is in the range [$35$-$45$]mW, both at 80~MHz repetition rate ($1$~mW corresponds to $12.5$~pJ pulse energy). We measured for different powers the Stokes intensity generated by the write pulse, as well as the anti-Stokes intensity generated by the read pulse (which is close in frequency to the write pulse, so no significant difference in Raman cross section is expected). The integrated Stokes and anti-Stokes counts are plotted in Figs.~\ref{fig:power}a and \ref{fig:power}b respectively.The power dependence is best well fit by a linear regression for both of them and shows no sign of quadratic increase, excluding a significant contribution of stimulated Raman scattering . 

In addition, these data put a stringent upper bound on the mean number of Stokes photons generated by pulse (in the spatial mode we collect). Indeed, for the anti-Stokes signal, the crossover from linear to quadratic power dependence occurs when the mean Stokes photon number (and therefore generated phonon number) per pulse is on the order of the thermal occupancy \cite{schmidt_linking_2017, schmidt_quantum_2016}, i.e. $0.04$. We find no sign of quadratic increase up to $100$~mW pump power. Therefore, in the conditions of the correlation measurements ($\sim5$~mW write power), the mean excited phonon number is expected to be on the order of $2\times10^{-3}$.

\end{document}